\documentclass[prd,aps,showpacs,epsf,floats,onecolumn]{revtex4}%
\usepackage{amssymb}
\usepackage{amsfonts}
\usepackage{amsmath}
\usepackage{graphicx}
\usepackage{color}%
\providecommand{\U}[1]{\protect\rule{.1in}{.1in}}
\begin{document}
\title{\textbf{Minimum time for the evolution to a nonorthogonal quantum state and
upper bound of the geometric efficiency of quantum evolutions}}
\author{\textbf{Carlo Cafaro}$^{1}$\textbf{ and Paul M.\ Alsing}$^{2}$}
\affiliation{$^{1}$SUNY Polytechnic Institute, 12203 Albany, New York, USA}
\affiliation{$^{2}$Air Force Research Laboratory, Information Directorate, 13441 Rome, New
York, USA}

\begin{abstract}
We present a simple proof of the fact that the minimum time $T_{AB}$ for the
quantum evolution between two arbitrary states $\left\vert A\right\rangle $
and $\left\vert B\right\rangle $ equals $T_{AB}=\hslash\cos^{-1}\left[
\left\vert \left\langle A|B\right\rangle \right\vert \right]  /\Delta E$ with
$\Delta E$ being the constant energy uncertainty of the system. This proof is
performed in the absence of any geometrical arguments. Then, being in the
geometric framework of quantum evolutions based upon the geometry of the
projective Hilbert space, we discuss the roles played by either minimum-time
or maximum-energy uncertainty concepts in defining a geometric efficiency
measure $\varepsilon$ of quantum evolutions between two arbitrary quantum
states. Finally, we provide a quantitative justification of the validity of
the inequality $\varepsilon\leq1$ even when the system passes only through
nonorthogonal quantum states.

\end{abstract}

\maketitle

\section{Introduction}

There are several works on the minimum time for the quantum evolution to an
orthogonal state. A first list of works on a lower bound on the orthogonality
time based on the energy spread includes the investigations by Mandelstam and
Tamm \cite{mandelstam45}, Fleming \cite{fleming73}, Anandan and Aharonov
\cite{anandan90}, and Vaidman \cite{vaidman92}. In a second type of
explorations, there are works like the one by Margolous and Levitin
\cite{margolus98} that express a lower bound on the orthogonalization time
based on the average energy of the system. In Ref. \cite{levitin09}, instead,
Levitin and Toffoli proposed a lower bound that involves both the energy
spread and the average energy of the system. These lower bounds have been
extended from isolated to non-isolated systems \cite{svozil05} and, moreover,
from pure to mixed quantum states in the presence of entanglement
\cite{giovannetti03,zander07} as well. For a recent expository review of the
minimum evolution time and quantum speed limit inequalities that includes
generalizations to mixed system states and open multiparticle systems, we
refer to Ref. \cite{frey16} and references therein.

It is known that the minimum time $T$ for the Schr\"{o}dinger evolution to an
orthogonal state for a system with a time-independent energy uncertainty
$\Delta E$ is given by $T=h/\left(  4\Delta E\right)  $ with $h$ denoting the
Planck constant. This finding was proved, for example, by Vaidman in Ref.
\cite{vaidman92} without using any geometric argument, while it was proved by
Anandan and Aharonov in Ref. \cite{anandan90} with the help of more elaborate
geometric considerations within the framework of the projective Hilbert space
geometry. In both Refs. \cite{vaidman92} and \cite{anandan90}, the expression
for the minimum time was restricted to evolutions to an orthogonal quantum
state. More specifically, Anandan and Aharonov derived a rather interesting
inequality relating the time interval $\Delta t$ of the quantum evolution to
the time-averaged uncertainty in energy $\left\langle \Delta E\right\rangle $
(during the time interval $\Delta t$) in Ref. \cite{anandan90},%
\begin{equation}
\left\langle \Delta E\right\rangle \Delta t\geq h/4\text{.} \label{inequality}%
\end{equation}
In particular, they stated that the equality sign in Eq. (\ref{inequality})
holds if and only if the system moves along a geodesic in the projective
Hilbert space. To quantify geodesic motion in the projective Hilbert space,
they also introduced a notion of efficiency denoted as $\varepsilon$ (this
symbol $\varepsilon$ will be replaced with $\eta_{\text{QM}}^{\left(
\mathrm{geometric}\right)  }$ in this manuscript), with $\varepsilon\leq1$
containing the inequality in Eq. (\ref{inequality}) as a special case.
However, they stated without proof that $\varepsilon\leq1$ is valid more
generally even when the system does not pass through orthogonal states.

In this paper, inspired by the results presented in Refs.
\cite{vaidman92,anandan90}, we find the expression for the minimum time for
the evolution to an arbitrary nonorthogonal quantum state. Moreover, based on
this first result, we provide a quantitative justification of the validity of
the inequality $\varepsilon\leq1$ even when the system passes only through
nonorthogonal states.

The layout of the remaining of the paper is as follows. In Section II, we
present a derivation of the expression for the minimum time for the evolution
to an arbitrary nonorthogonal quantum state. In Section III, exploiting the
result obtained in Section II, we provide a quantitative justification of the
validity of the inequality $\varepsilon\leq1$ even when the system passes only
through nonorthogonal states. In Section IV, we discuss in an explicit manner
the concepts of minimum evolution time and quantum geometric efficiency in two
physical examples. Finally, our final considerations appear in Section V.

\section{Minimum time without geometric arguments}

In this Section, we provide a proof that the minimum time $T_{AB}$ for the
quantum evolution between two arbitrary states $\left\vert A\right\rangle $
and $\left\vert B\right\rangle $ is equal to $T_{AB}=\hslash\cos^{-1}\left[
\left\vert \left\langle A|B\right\rangle \right\vert \right]  /\Delta E$ where
$\Delta E$ denotes the constant energy uncertainty of the system. This proof
follows closely the proof presented by Vaidman in Ref. \cite{vaidman92} where,
however, the quantum evolution was restricted to an orthogonal quantum state.

We begin with some preliminary remarks. Consider an operator $\hat{Q}$ and a
normalized quantum state $\left\vert \psi\right\rangle $. The state $\hat{Q}$
$\left\vert \psi\right\rangle $ can be decomposed as,%
\begin{equation}
\hat{Q}\left\vert \psi\right\rangle =\alpha_{1}\left\vert \psi\right\rangle
+\alpha_{2}\left\vert \psi_{\bot}\right\rangle \text{,}%
\end{equation}
where $\alpha_{1}$, $\alpha_{2}\in%
\mathbb{C}
$ and $\left\vert \psi\right\rangle \perp\left\vert \psi_{\bot}\right\rangle
$. Then, we get%
\begin{align}
\hat{Q}\left\vert \psi\right\rangle  &  =\alpha_{1}\left\vert \psi
\right\rangle +\alpha_{2}\left\vert \psi_{\bot}\right\rangle \nonumber\\
&  =\left\vert \alpha_{1}\right\vert e^{i\varphi_{1}}\left\vert \psi
\right\rangle +\left\vert \alpha_{2}\right\vert e^{i\varphi_{2}}\left\vert
\psi_{\bot}\right\rangle \nonumber\\
&  =e^{i\varphi_{2}}\left[  \left\vert \alpha_{1}\right\vert e^{i\left(
\varphi_{1}-\varphi_{2}\right)  }\left\vert \psi\right\rangle +\left\vert
\alpha_{2}\right\vert \left\vert \psi_{\bot}\right\rangle \right]
\text{.}\label{t1a}%
\end{align}
The state $\hat{Q}\left\vert \psi\right\rangle $ in Eq. (\ref{t1a}) is
physically equivalent to the state $\left\vert \alpha_{1}\right\vert
e^{-i\varphi}\left\vert \psi\right\rangle +\left\vert \alpha_{2}\right\vert
\left\vert \psi_{\bot}\right\rangle $ with $\varphi\overset{\text{def}}%
{=}\varphi_{2}-\varphi_{1}$ given that global\textbf{ }phases have no
relevance in quantum mechanics. Therefore, we conclude from Eq. (\ref{t1a})
that we can write the state $\hat{Q}\left\vert \psi\right\rangle $ as%
\begin{equation}
\hat{Q}\left\vert \psi\right\rangle =\alpha\left\vert \psi\right\rangle
+\beta\left\vert \psi_{\bot}\right\rangle \text{,}\label{t4}%
\end{equation}
with $\alpha\in%
\mathbb{C}
$ and $\beta\in%
\mathbb{R}
_{+}$. If we assume that $\hat{Q}$ is also Hermitian, we have%
\begin{equation}
\alpha=\left\langle \psi|\hat{Q}|\psi\right\rangle =\left\langle \psi|\hat
{Q}^{\dagger}|\psi\right\rangle =\left\langle \psi|\hat{Q}|\psi\right\rangle
^{\ast}=\alpha^{\ast}\text{,}\label{t2}%
\end{equation}
that is, $\alpha\in%
\mathbb{R}
$. Moreover, remaining in the working condition with $\hat{Q}$ being
Hermitian, the dispersion $\Delta Q^{2}$ of the operator $\hat{Q}$ becomes%
\begin{align}
\Delta Q^{2} &  =\left\langle \psi|\hat{Q}^{2}|\psi\right\rangle -\left\langle
\psi|\hat{Q}|\psi\right\rangle ^{2}\nonumber\\
&  =\left\langle \psi|\hat{Q}\hat{Q}^{\dagger}|\psi\right\rangle -\alpha
^{2}\nonumber\\
&  =\left\vert \alpha\right\vert ^{2}+\left\vert \beta\right\vert ^{2}%
-\alpha^{2}\nonumber\\
&  =\alpha^{2}+\beta^{2}-\alpha^{2}\nonumber\\
&  =\beta^{2}\text{,}%
\end{align}
that is,%
\begin{equation}
\beta=\Delta Q\text{.}\label{t3}%
\end{equation}
Finally, using Eqs. (\ref{t2}) and (\ref{t3}), $\hat{Q}\left\vert
\psi\right\rangle $ in Eq. (\ref{t4}) can be decomposed as%
\begin{equation}
\hat{Q}\left\vert \psi\right\rangle =\left\langle \hat{Q}\right\rangle
\left\vert \psi\right\rangle +\Delta Q\left\vert \psi_{\bot}\right\rangle
\text{,}\label{deco}%
\end{equation}
with $\left\langle \psi|\psi_{\bot}\right\rangle =0$. For an alternative
derivation of Eq. (\ref{deco}), we refer to Appendix A. Eq. (\ref{deco}) will
play a key role in our derivation of the minimum time expression.

At this point, we state the problem we wish to address. We want to find an
explicit expression of the minimum time for the evolution from the normalized
state $\left\vert A\right\rangle =\left\vert \psi\left(  0\right)
\right\rangle $ to the state $\left\vert B\right\rangle =\left\vert
\psi\left(  T_{AB}\right)  \right\rangle $ with $\left\langle A|B\right\rangle
\neq0$ in the working condition that the dispersion of the Hamiltonian
operator $\mathrm{\hat{H}}$ is constant,%
\begin{equation}
\Delta E=\sqrt{\left\langle A|\mathrm{\hat{H}}^{2}|A\right\rangle
-\left\langle A|\mathrm{\hat{H}}|A\right\rangle ^{2}}=\text{constant.}
\label{dispersion}%
\end{equation}
\ 

Before continuing our proof, we would like to emphasize at this point that our
demonstration works equally well for time-dependent Hamiltonians\textbf{
}$\mathrm{\hat{H}}\left(  t\right)  $\textbf{ }with non constant energy
uncertainty\textbf{ }$\Delta E\left(  t\right)  $ where $\Delta E^{2}\left(
t\right)  \overset{\text{def}}{=}\left\langle \psi\left(  t\right)
|\mathrm{\hat{H}}^{2}|\psi\left(  t\right)  \right\rangle -\left\langle
\psi\left(  t\right)  |\mathrm{\hat{H}}|\psi\left(  t\right)  \right\rangle
^{2}$\textbf{ }and\textbf{ }$\left\vert \psi\left(  t\right)  \right\rangle
$\textbf{ }is assumed to be normalized to one. However, as we shall see, an
expression of the minimum evolution time can be only obtained in an implicit
manner when taking into consideration systems specified by this type of Hamiltonians.

Returning to our proof, we point out for the sake of completeness that since
stationary states are quantum states with no energy uncertainty (that is,
$\Delta E=0$ \cite{merzbacker}), we shall limit our considerations to quantum
evolutions of nonstationary states \cite{fleming73}. Following the work by
Vaidman \cite{vaidman92,vaidman90}, the minimum time of the evolution to a
nonorthogonal state can be found by evaluating the maximum of the absolute
value of the rate of change of the modulus squared of the quantum overlap
$\left\langle \psi\left(  t\right)  |A\right\rangle $ with $\left\vert
\psi\left(  t\right)  \right\rangle $ being an intermediate state between
$\left\vert A\right\rangle $ and $\left\vert B\right\rangle $ satisfying the
time-dependent Schr\"{o}dinger equation,
\begin{equation}
i\hslash\partial_{t}\left\vert \psi\left(  t\right)  \right\rangle
=\mathrm{\hat{H}}\left\vert \psi\left(  t\right)  \right\rangle \text{.}%
\label{deco1}%
\end{equation}
Let us consider the quantity $d\left[  \left\vert \left\langle \psi\left(
t\right)  |A\right\rangle \right\vert ^{2}\right]  /dt$. Observe that,%
\begin{equation}
\left\vert \left\langle \psi\left(  t\right)  |A\right\rangle \right\vert
^{2}=\left\langle \psi\left(  t\right)  |A\right\rangle \left\langle
\psi\left(  t\right)  |A\right\rangle ^{\ast}=\left\langle \psi\left(
t\right)  |A\right\rangle \left\langle A|\psi\left(  t\right)  \right\rangle
\text{.}%
\end{equation}
Therefore, the rate of change in time of the modulus squared of the quantum
overlap $\left\langle \psi\left(  t\right)  |A\right\rangle $ becomes%
\begin{align}
\frac{d\left[  \left\vert \left\langle \psi\left(  t\right)  |A\right\rangle
\right\vert ^{2}\right]  }{dt} &  =\left\langle \dot{\psi}|A\right\rangle
\left\langle A|\psi\right\rangle +\left\langle \psi|A\right\rangle
\left\langle A|\dot{\psi}\right\rangle \nonumber\\
&  =\left\langle \psi|A\right\rangle \left\langle A|\dot{\psi}\right\rangle
+\left\langle \psi|A\right\rangle ^{\ast}\left\langle A|\dot{\psi
}\right\rangle ^{\ast}\nonumber\\
&  =2\operatorname{Re}\left[  \left\langle \psi|A\right\rangle \left\langle
A|\dot{\psi}\right\rangle \right]  \text{,}\label{de2}%
\end{align}
where $\left\vert \dot{\psi}\right\rangle =\left\vert \frac{d\psi}%
{dt}\right\rangle =\frac{d\left\vert \psi\right\rangle }{dt}$. Using Eqs.
(\ref{deco}) with $\hat{Q}=\mathrm{\hat{H}}$ together with Eq. (\ref{deco1}),
we obtain%
\begin{equation}
\left\vert \dot{\psi}\right\rangle =-\frac{i}{\hslash}\mathrm{\hat{H}%
}\left\vert \psi\right\rangle =-\frac{i}{\hslash}\left[  \left\langle
\mathrm{\hat{H}}\right\rangle \left\vert \psi\right\rangle +\Delta E\left\vert
\psi_{\bot}\right\rangle \right]  \text{.}\label{de3}%
\end{equation}
Using Eq. (\ref{de3}), Eq. (\ref{de2}) becomes%
\begin{align}
\frac{d\left[  \left\vert \left\langle \psi\left(  t\right)  |A\right\rangle
\right\vert ^{2}\right]  }{dt} &  =2\operatorname{Re}\left\{  \left\langle
\psi|A\right\rangle \cdot\left(  -\frac{i}{\hslash}\right)  \cdot\left[
\left\langle \mathrm{\hat{H}}\right\rangle \left\langle A|\psi\right\rangle
+\Delta E\left\langle A|\psi_{\bot}\right\rangle \right]  \right\}
\nonumber\\
&  =2\operatorname{Re}\left[  -\frac{i}{\hslash}\left\langle \mathrm{\hat{H}%
}\right\rangle \left\vert \left\langle \psi|A\right\rangle \right\vert
^{2}-\frac{i}{\hslash}\Delta E\left\langle \psi|A\right\rangle \left\langle
A|\psi_{\bot}\right\rangle \right]  \nonumber\\
&  =-2\frac{\Delta E}{\hslash}\operatorname{Re}\left[  i\left\langle
\psi|A\right\rangle \left\langle A|\psi_{\bot}\right\rangle \right]  \text{,}%
\end{align}
that is,%
\begin{equation}
\frac{d\left[  \left\vert \left\langle \psi\left(  t\right)  |A\right\rangle
\right\vert ^{2}\right]  }{dt}=-2\frac{\Delta E}{\hslash}\operatorname{Re}%
\left[  i\left\langle \psi|A\right\rangle \left\langle A|\psi_{\bot
}\right\rangle \right]  \text{.}\label{de4}%
\end{equation}
For a given value of $\Delta E$ and $\left\vert \left\langle \psi
|A\right\rangle \right\vert $, the absolute value of the RHS of Eq.
(\ref{de4}) achieves its maximum number when $\left\vert \left\langle
A|\psi_{\bot}\right\rangle \right\vert $ is maximum by observing that%
\begin{equation}
\left\vert \operatorname{Re}\left[  i\left\langle \psi|A\right\rangle
\left\langle A|\psi_{\bot}\right\rangle \right]  \right\vert \leq\left\vert
i\left\langle \psi|A\right\rangle \left\langle A|\psi_{\bot}\right\rangle
\right\vert \leq\left\vert \left\langle \psi|A\right\rangle \right\vert
\left\vert \left\langle A|\psi_{\bot}\right\rangle \right\vert \text{.}%
\end{equation}
To find the maximum of $\left\vert \left\langle A|\psi_{\bot}\right\rangle
\right\vert $, we proceed as follows. In general, the resolution of the
identity for an $n$-dimensional Hilbert space $\mathcal{H}$ is given by%
\begin{equation}
\mathrm{\hat{1}}=%
{\displaystyle\sum\limits_{i=1}^{n}}
\left\vert \psi_{i}\right\rangle \left\langle \psi_{i}\right\vert
\text{,}\label{general}%
\end{equation}
with $\left\langle \psi_{i}|\psi_{j}\right\rangle =\delta_{ij}$ for any $1\leq
i$, $j\leq n$. For clarity of exposition and without loss of generality, we
assume here a \ resolution of the identity on the full Hilbert space
$\mathcal{H}$\textbf{ }that can be recast as $\mathrm{\hat{1}}\overset
{\text{def}}{=}\left\vert \psi\right\rangle \left\langle \psi\right\vert
+\left\vert \psi_{\bot}\right\rangle \left\langle \psi_{\bot}\right\vert
+\left\vert \psi_{\bot\bot}\right\rangle \left\langle \psi_{\bot\perp
}\right\vert $. We note that
\begin{align}
\left\vert A\right\rangle  &  =\mathrm{\hat{1}}\left\vert A\right\rangle
\nonumber\\
&  =\left[  \left\vert \psi\right\rangle \left\langle \psi\right\vert
+\left\vert \psi_{\bot}\right\rangle \left\langle \psi_{\bot}\right\vert
+\left\vert \psi_{\bot\bot}\right\rangle \left\langle \psi_{\bot\perp
}\right\vert \right]  \left\vert A\right\rangle \nonumber\\
&  =\left\langle \psi|A\right\rangle \left\vert \psi\right\rangle
+\left\langle \psi_{\bot}|A\right\rangle \left\vert \psi_{\bot}\right\rangle
+\left\langle \psi_{\bot\bot}|A\right\rangle \left\vert \psi_{\bot\bot
}\right\rangle \text{,}\label{de5}%
\end{align}
where,%
\begin{equation}
\left\langle \psi\left(  t\right)  |\psi_{\bot}\left(  t\right)  \right\rangle
=\left\langle \psi\left(  t\right)  |\psi_{\bot\bot}\left(  t\right)
\right\rangle =\left\langle \psi_{\bot}\left(  t\right)  |\psi_{\bot\bot
}\left(  t\right)  \right\rangle =0\text{, }\forall t\text{.}\label{de6}%
\end{equation}
Since $\left\langle A|A\right\rangle =1$, from Eqs. (\ref{de5}) and
(\ref{de6}) we obtain%
\begin{equation}
\left\vert \left\langle \psi|A\right\rangle \right\vert ^{2}+\left\vert
\left\langle \psi_{\bot}|A\right\rangle \right\vert ^{2}+\left\vert
\left\langle \psi_{\bot\bot}|A\right\rangle \right\vert ^{2}=1\text{.}%
\label{t6}%
\end{equation}
Therefore, we get from Eq. (\ref{t6}) that%
\begin{equation}
\left\vert \left\langle A|\psi_{\bot}\right\rangle \right\vert ^{2}%
=1-\left\vert \left\langle \psi|A\right\rangle \right\vert ^{2}-\left\vert
\left\langle \psi_{\bot\bot}|A\right\rangle \right\vert ^{2}\text{,}%
\end{equation}
that is, $\left\vert \left\langle A|\psi_{\bot}\right\rangle \right\vert $ is
maximum when $\left\langle \psi_{\bot\bot}|A\right\rangle =0$. Its maximum
value equals,%
\begin{equation}
\left\vert \left\langle A|\psi_{\bot}\right\rangle \right\vert _{\max}%
=\sqrt{1-\left\vert \left\langle A|\psi\right\rangle \right\vert ^{2}}%
\text{.}\label{max}%
\end{equation}

For completeness, we note that in the general scenario where one employs Eq.
(\ref{general}), $\left\vert \left\langle A|\psi_{\bot}\right\rangle
\right\vert $ is maximum when $\left\langle \psi_{i}|A\right\rangle =0$ for
any $i=3$,..., $n$ with $\left\vert \psi_{1}\right\rangle $ and $\left\vert
\psi_{2}\right\rangle $ corresponding to $\left\vert \psi\right\rangle $ and
$\left\vert \psi_{\perp}\right\rangle $, respectively. 

Before continuing our proof, we would like to remark at this point that it is
straightforward to see that the subscript\textbf{ \textquotedblleft}$\max
$\textbf{\textquotedblright\ }in\ Eq. (\ref{max}) is not strictly necessary if
we employ a different resolution of the identity operator $\mathrm{\hat{1}}$
on the full Hilbert space with orthogonal decomposition given by
$\mathcal{H}\overset{\text{def}}{=}\mathcal{H}_{\psi}\oplus\mathcal{H}%
_{\psi_{\bot}}$ as used in Appendix A, and not by $\mathcal{H}\overset
{\text{def}}{=}\mathcal{H}_{\psi}\oplus\mathcal{H}_{\psi_{\bot}}%
\oplus\mathcal{H}_{\psi_{\bot\bot}}$ as in Eq.\textbf{ }(\ref{de5}).

Returning to our proof, we note from Eqs. (\ref{de4}) and (\ref{max}) that the
maximum of the absolute value of the rate of change of the quantum overlap
depends only on $\Delta E$ and $\left\vert \left\langle A|\psi\right\rangle
\right\vert $ and is given by%
\begin{equation}
\left\vert \frac{d\left[  \left\vert \left\langle \psi\left(  t\right)
|A\right\rangle \right\vert ^{2}\right]  }{dt}\right\vert _{\max}%
=2\frac{\Delta E}{\hslash}\left\vert \left\langle A|\psi\right\rangle
\right\vert \sqrt{1-\left\vert \left\langle A|\psi\right\rangle \right\vert
^{2}}\text{.}%
\end{equation}
Finally, to get the fastest evolution to a nonorthogonal state, we impose%
\begin{equation}
\frac{d\left[  \left\vert \left\langle \psi|A\right\rangle \right\vert
^{2}\right]  }{dt}=-2\frac{\Delta E}{\hslash}\left\vert \left\langle
A|\psi\right\rangle \right\vert \sqrt{1-\left\vert \left\langle A|\psi
\right\rangle \right\vert ^{2}}\text{.} \label{du7}%
\end{equation}
For the sake of completeness, we point out that this constraint entails
imposing a phase relationship between\textbf{ }$\left\langle \psi
|A\right\rangle =|\left\langle \psi|A\right\rangle |\,e^{i\,\varphi_{\psi}}%
$\textbf{ }and $\left\langle \psi_{\perp}|A\right\rangle =|\left\langle
\psi_{\perp}|A\right\rangle |\,e^{i\,\varphi_{\psi_{\perp}}}$ such that
$\varphi_{\psi_{\perp}}-\varphi_{\psi}=\pi/2$. Letting\textbf{ }$\left\vert
\left\langle \psi|A\right\rangle \right\vert \overset{\text{def}}{=}%
\cos\left(  \theta\right)  $, Eq. (\ref{du7}) yields%
\begin{equation}
\frac{d\left[  \cos^{2}\left(  \theta\right)  \right]  }{dt}=-2\frac{\Delta
E}{\hslash}\cos\left(  \theta\right)  \sin\left(  \theta\right)  \text{,}%
\end{equation}
that is,%
\begin{equation}
-2\cos\left(  \theta\right)  \sin\left(  \theta\right)  \dot{\theta}%
=-2\frac{\Delta E}{\hslash}\cos\left(  \theta\right)  \sin\left(
\theta\right)  \text{.} \label{t7}%
\end{equation}
Thus, we get from Eq. (\ref{t7}) that%
\begin{equation}
\dot{\theta}=\frac{\Delta E}{\hslash}\text{.} \label{du8}%
\end{equation}
Integrating Eq. (\ref{du8}), we obtain%
\begin{equation}
\int_{\theta\left(  0\right)  }^{\theta\left(  T_{AB}\right)  }d\theta
=\int_{0}^{T_{AB}}\frac{\Delta E}{\hslash}dt\text{.} \label{du9}%
\end{equation}
Recalling that $\left\vert \left\langle \psi\left(  t\right)  |A\right\rangle
\right\vert \overset{\text{def}}{=}\cos\left[  \theta\left(  t\right)
\right]  $ with\textbf{ }$\theta\left(  0\right)  =0$ since $\left\vert
\psi\left(  0\right)  \right\rangle \overset{\text{def}}{=}\left\vert
A\right\rangle $, after some simple algebra, we finally get from Eq.
(\ref{du9})%
\begin{equation}
T_{AB}=\frac{\hslash}{\Delta E}\cos^{-1}\left[  \left\vert \left\langle
A|B\right\rangle \right\vert \right]  \text{,} \label{du10}%
\end{equation}
that is,%
\begin{equation}
\Delta E\cdot T_{AB}=\hslash\cos^{-1}\left[  \left\vert \left\langle
A|B\right\rangle \right\vert \right]  \text{.}%
\end{equation}
The quantity $T_{AB}$ in Eq. (\ref{du10})\ denotes the minimum time interval
needed for the evolution (unitary Schr\"{o}dinger evolution with the
assumption of constant dispersion of the Hamiltonian operator) from
$\left\vert A\right\rangle $ to $\left\vert B\right\rangle $ with the two
states being nonorthogonal. As a side remark, we point out that when
$\left\vert A\right\rangle $ and $\left\vert B\right\rangle $ are orthogonal,
Eq. (\ref{du10}) yields%
\begin{equation}
\Delta E\cdot T_{AB}^{\perp}=\frac{h}{4}\text{.} \label{t8}%
\end{equation}
Eq. (\ref{t8}) is the result that was originally obtained,without use of
geometrical reasoning, by Vaidman in Ref. \cite{vaidman92}. As a final remark,
we point out that Eq. (\ref{du10}) can be recast as%
\begin{equation}
T_{AB}=\frac{\hslash}{\Delta E}\sin^{-1}\left[  \sqrt{1-\left\vert
\left\langle A|B\right\rangle \right\vert ^{2}}\right]  \text{.} \label{du11}%
\end{equation}
Eq. (\ref{du11}) reduces to the optimal time expression obtained by Bender and
collaborators in Ref. \cite{bender07} when setting $\left\vert A\right\rangle
\overset{\text{def}}{=}\left\vert 0\right\rangle $ and $\left\vert
B\right\rangle \overset{\text{def}}{=}a\left\vert 0\right\rangle +b\left\vert
1\right\rangle $ with $a$, $b\in%
\mathbb{C}
$.

As pointed out right below Eq. (\ref{dispersion}), our demonstration leading
to Eq. (\ref{du9}) works equally well for time-dependent Hamiltonians\textbf{
}$\mathrm{\hat{H}}\left(  t\right)  $. In particular, Eq. (\ref{du9}) remains
valid if we replace a constant $\Delta E$\textbf{ }with a time-dependent
$\Delta E\left(  t\right)  $. In this case, integration of Eq. (\ref{du9})
yields
\begin{equation}
\left\langle \Delta E\right\rangle _{T_{AB}}T_{AB}=\hslash\cos^{-1}\left[
\left\vert \left\langle A|B\right\rangle \right\vert \right]  \text{,}
\label{minimi}%
\end{equation}
where $\left\langle \Delta E\right\rangle _{T_{AB}}$ is the time-averaged
uncertainty during the time interval\textbf{ }$T_{AB}$ defined as,
\begin{equation}
\left\langle \Delta E\right\rangle _{T_{AB}}\overset{\text{def}}{=}\frac
{1}{T_{AB}}\int_{0}^{T_{AB}}\Delta E\left(  t^{\prime}\right)  dt^{\prime
}=\frac{\mathcal{E}\left(  T_{AB}\right)  -\mathcal{E}\left(  0\right)
}{T_{AB}}\text{,} \label{minimi1}%
\end{equation}
with $d\mathcal{E}/dt^{\prime}\overset{\text{def}}{=}\Delta E\left(
t^{\prime}\right)  $. It is clear from Eqs. (\ref{minimi}) and (\ref{minimi1})
that\textbf{ \ }$\mathcal{E}\left(  T_{AB}\right)  =\mathcal{E}\left(
0\right)  +\hslash\cos^{-1}\left[  \left\vert \left\langle A|B\right\rangle
\right\vert \right]  $ and, therefore, we expect that a closed form analytical
expression of the minimum evolution time\textbf{ }$T_{AB}$\textbf{ }cannot be
generally obtained in an explicit manner in such more realistic time-dependent scenarios.

\section{Efficiency with geometric arguments}

Recall that in the geometric formulation of quantum mechanical
Schr\"{o}dinger's evolution, one can consider a measure of efficiency that
quantifies the departure of an effective (non-geodesic evolution, in general)
from an ideal geodesic evolution. Such a geodesic evolution is characterized
by paths of shortest length that connect initial and final quantum states
$\left\vert A\right\rangle $ and $\left\vert B\right\rangle $, respectively.
In particular, under this scheme one can define an efficiency in geometric
quantum mechanics that takes into account a quantum mechanical evolution of a
state vector $\left\vert \psi\left(  t\right)  \right\rangle $ in an
$N$-dimensional complex Hilbert space specified by Schr\"{o}dinger equation,%
\begin{equation}
i\hslash\partial_{t}\left\vert \psi\left(  t\right)  \right\rangle
=\mathrm{\hat{H}}\left(  t\right)  \left\vert \psi\left(  t\right)
\right\rangle \text{,}%
\end{equation}
with $0\leq t\leq T_{AB}$, $\hslash\overset{\text{def}}{=}h/\left(
2\pi\right)  $, $h$ being the Planck constant, and $\mathrm{\hat{H}}$ denoting
the Hamiltonian of the system. A geometric measure of efficiency
$\eta_{\text{QM}}^{\left(  \mathrm{geometric}\right)  }$ with $0\leq
\eta_{\text{QM}}^{\left(  \mathrm{geometric}\right)  }\leq1$ for such a
quantum system can be defined as \cite{anandan90,cafaro20A},%
\begin{equation}
\eta_{\text{QM}}^{\left(  \mathrm{geometric}\right)  }\overset{\text{def}}%
{=}\frac{s_{0}}{s}=1-\frac{\Delta s}{s}=\frac{2\cos^{-1}\left[  \left\vert
\left\langle \psi\left(  0\right)  |\psi\left(  T_{AB}\right)  \right\rangle
\right\vert \right]  }{2\int_{0}^{T_{AB}}\frac{\Delta E\left(  t^{\prime
}\right)  }{\hslash}dt^{\prime}}\text{,} \label{effi}%
\end{equation}
where $\Delta s\overset{\text{def}}{=}s-s_{0}$, $s_{0}\overset{\text{def}}%
{=}2\cos^{-1}\left[  \left\vert \left\langle \psi\left(  0\right)
|\psi\left(  T_{AB}\right)  \right\rangle \right\vert \right]  $ denotes the
distance along the (ideal) shortest geodesic path joining the distinct initial
$\left\vert \psi\left(  0\right)  \right\rangle \overset{\text{def}}{=}$
$\left\vert A\right\rangle $ and final $\left\vert \psi\left(  T_{AB}\right)
\right\rangle \overset{\text{def}}{=}\left\vert B\right\rangle $ states on the
projective Hilbert space $%
\mathbb{C}
P^{N-1}$ and finally, $s\overset{\text{def}}{=}2\int_{0}^{T_{AB}}\left[
\Delta E\left(  t^{\prime}\right)  \right]  /\hslash dt^{\prime}$ is the
distance along the (real) actual dynamical trajectory traced by the state
vector $\left\vert \psi\left(  t\right)  \right\rangle $ with $0\leq t\leq
T_{AB}$ and finally, $\Delta E$ represents the uncertainty in the energy of
the system. We emphasize that the numerator in\ Eq. (\ref{effi}) is the angle
between the state vectors $\left\vert \psi\left(  0\right)  \right\rangle $
and $\left\vert \psi\left(  T_{AB}\right)  \right\rangle $ and is equal to the
Wootters distance $ds_{\text{Wootters}}$ \cite{wootters81},%
\begin{equation}
ds_{\text{Wootters}}\left(  \left\vert \psi\left(  0\right)  \right\rangle
\text{, }\left\vert \psi\left(  T_{AB}\right)  \right\rangle \right)
\overset{\text{def}}{=}2\cos^{-1}\left[  \left\vert \left\langle \psi\left(
0\right)  |\psi\left(  T_{AB}\right)  \right\rangle \right\vert \right]
\text{.}%
\end{equation}
Furthermore, the denominator in Eq. (\ref{effi}) represents the integral of
the infinitesimal distance $ds$ along the evolution curve in the projective
Hilbert space \cite{anandan90},%
\begin{equation}
ds\overset{\text{def}}{=}2\frac{\Delta E\left(  t\right)  }{\hslash}dt\text{.}
\label{distance}%
\end{equation}
Curiously, Anandan and Aharonov proved that the infinitesimal distance $ds$ in
Eq. (\ref{distance}) is connected to the Fubini-Study infinitesimal distance
$ds_{\text{Fubini-Study}}$ by the following condition,%
\begin{equation}
ds_{\text{Fubini-Study}}^{2}\left(  \left\vert \psi\left(  t\right)
\right\rangle \text{, }\left\vert \psi\left(  t+dt\right)  \right\rangle
\right)  \overset{\text{def}}{=}4\left[  1-\left\vert \left\langle \psi\left(
t\right)  |\psi\left(  t+dt\right)  \right\rangle \right\vert ^{2}\right]
=4\frac{\Delta E^{2}\left(  t\right)  }{\hslash^{2}}dt^{2}+\mathcal{O}\left(
dt^{3}\right)  \text{,} \label{relation}%
\end{equation}
with $\mathcal{O}\left(  dt^{3}\right)  $ denoting an infinitesimal quantity
equal or higher than $dt^{3}$. Eqs. (\ref{distance}) and (\ref{relation})
imply that $s$ is proportional to the time integral of the uncertainty in
energy $\Delta E$ of the system and specifies the distance along the quantum
evolution of the system in the projective Hilbert space as measured by the
Fubini-Study metric. We point out that when the actual dynamical curve
coincides with the shortest geodesic path connecting the initial and final
states, $\Delta s$ is equal to zero and the efficiency $\eta_{\text{QM}%
}^{\left(  \mathrm{geometric}\right)  }$ in Eq. (\ref{effi}) equals one.
Obviously, $\pi$ is the shortest possible distance between two orthogonal
states in the projective Hilbert space. In general, however, $s\geq\pi$ for
such a pair of orthogonal pure states.

Given the important role played by energy uncertainty in the geometry of
quantum evolutions, one may wonder whether or not there is some sort of
quantum mechanical uncertainty relation in this geometric framework. We recall
that the standard quantum mechanical uncertainty relation is given by
\cite{peres95},%
\begin{equation}
\Delta x\Delta p\geq\hslash/2\text{.} \label{heisenberg}%
\end{equation}
Eq. (\ref{heisenberg}) mirrors the intrinsic randomness of the outcomes of
quantum experiments.\ Precisely, if one repeats several times the same state
preparation scheme and then measures the operators $x$ or $p$, the
observations collected for $x$ and $p$ are specified by standard deviations
$\Delta x$ and $\Delta p$ whose product $\Delta x\Delta p$ is greater than
$\hslash/2$. Gaussian wave packets, in particular, are characterized by a
minimum position-momentum uncertainty defined by $\Delta x\Delta p=\hslash/2$.
In the geometry of quantum evolutions, there is an analogue of Eq.
(\ref{heisenberg}) where, for instance, Gaussian wave packets are replaced by
geodesic paths in the projective Hilbert space. Indeed, take into
consideration the time-averaged uncertainty in energy $\left\langle \Delta
E\right\rangle _{T_{AB}^{\bot}}$ during a time interval $T_{AB}^{\bot}$
defined as \cite{anandan90},%
\begin{equation}
\left\langle \Delta E\right\rangle _{T_{AB}^{\bot}}\overset{\text{def}}%
{=}\frac{1}{T_{AB}^{\bot}}\int_{0}^{T_{AB}^{\bot}}\Delta E\left(  t^{\prime
}\right)  dt^{\prime}\text{.} \label{energy}%
\end{equation}
The quantity $T_{AB}^{\bot}$ in Eq. (\ref{energy}) defines the
orthogonalization time, that is, the time interval during which the system
passes from an initial state $\left\vert A\right\rangle \overset{\text{def}%
}{=}\left\vert \psi\left(  0\right)  \right\rangle $ to a final state
$\left\vert B\right\rangle \overset{\text{def}}{=}\left\vert \psi\left(
T_{AB}^{\bot}\right)  \right\rangle $ where $\left\langle B|A\right\rangle
=\delta_{AB}$. Employing Eqs. (\ref{distance}) and (\ref{energy}) and
remembering that the shortest possible distance between two orthogonal quantum
states in the projective Hilbert space is $\pi$, we obtain%
\begin{equation}
\left\langle \Delta E\right\rangle _{T_{AB}^{\bot}}T_{AB}^{\bot}\geq
h/4\text{.} \label{uncertainty}%
\end{equation}
Specifically, the equality in Eq. (\ref{uncertainty}) holds only when the
quantum evolution is a geodesic evolution. Therefore, geodesic paths represent
minimum time-averaged energy uncertainty trajectories just as Gaussian wave
packets specify minimum position-momentum uncertainty wave packets.
Summarizing, when a quantum evolution exhibits minimum uncertainty
$\left\langle \Delta E\right\rangle _{T_{AB}^{\bot}}T_{AB}^{\bot}=h/4$, unit
efficiency $\eta_{\text{QM}}^{\left(  \mathrm{geometric}\right)  }=1$ is
achieved. This, in turn, occurs only if the physical systems moves along a
geodesic path in the projective Hilbert space. Interestingly, the
Anandan-Aharonov time-energy uncertainty relation in Eq. (\ref{uncertainty})
is related to the statistical speed of evolution\textbf{\ }$ds_{\text{FS}}%
/dt$\textbf{\ }of the physical system with\textbf{\ }$ds_{\text{FS}}^{2}%
$\textbf{\ }being the Fubini-Study infinitesimal line element squared.
Precisely, since\textbf{\ }$ds_{\text{FS}}/dt$ is proportional to $\Delta E$,
the system moves rapidly wherever the uncertainty in energy assumes large
values \cite{karol}.

We observe that if $\left\vert A\right\rangle $ and $\left\vert B\right\rangle
$ are orthogonal and, in addition, $\Delta E$ is constant, we have from Eq.
(\ref{effi}) that%
\begin{equation}
\eta_{\text{QM}}^{\left(  \mathrm{geometric}\right)  }=\frac{2\frac{\pi}{2}%
}{2\frac{\Delta E}{\hslash}T_{AB}^{\perp}}=\frac{h}{4}\frac{1}{\Delta E\cdot
T_{AB}^{\perp}}\text{.} \label{effi1}%
\end{equation}
Therefore, from Eq. (\ref{effi1}), we get%
\begin{equation}
\eta_{\text{QM}}^{\left(  \mathrm{geometric}\right)  }\leq1\Leftrightarrow
\frac{h}{4}\frac{1}{\Delta E\cdot T_{AB}^{\perp}}\leq1\Leftrightarrow\Delta
E\cdot T_{AB}^{\perp}\geq\frac{h}{4}\text{.} \label{c1}%
\end{equation}
In particular, we obtain%
\begin{equation}
\text{geodesic motion}\Leftrightarrow\eta_{\text{QM}}^{\left(
\mathrm{geometric}\right)  }=1\Leftrightarrow\frac{h}{4}\frac{1}{\Delta E\cdot
T_{AB}^{\perp}}=1\Leftrightarrow\Delta E\cdot T_{AB}^{\perp}=\frac{h}%
{4}\text{,}%
\end{equation}
that is,%
\begin{equation}
\text{\textrm{geodesic motion}}\Leftrightarrow\Delta E\cdot T_{AB}^{\perp
}=\frac{h}{4}\text{.} \label{gc1}%
\end{equation}
We remark that the geodesic constraint in Eq. (\ref{gc1}) is not a time-energy
uncertainty condition. Instead, it simply states that $\mathrm{distance}%
=\mathrm{speed}\times\mathrm{time}$. Indeed, using the Anandan-Aharonov
relation that states that the (angular) speed $v$ of a unitary evolution is
proportional to the energy uncertainty $\Delta E$ \cite{anandan90,cafaro20B},
$v=\left(  2\Delta E\right)  /\hslash$, the condition $\Delta E\cdot
T_{AB}^{\perp}=h/4$ can be recast as $s_{0}=vT_{AB}^{\perp}$ with
$s_{0}\overset{\text{def}}{=}\pi$. Furthermore, if $\left\vert A\right\rangle
$ and $\left\vert B\right\rangle $ are nonorthogonal and, in addition, $\Delta
E$ is constant, we have from Eq. (\ref{effi}) that%
\begin{equation}
\eta_{\text{QM}}^{\left(  \mathrm{geometric}\right)  }=\frac{\hslash}{\Delta
E\cdot T_{AB}}\cos^{-1}\left[  \left\vert \left\langle A|B\right\rangle
\right\vert \right]  \text{.} \label{effi2}%
\end{equation}
Therefore, Eq. (\ref{effi2}) yields%
\begin{equation}
\eta_{\text{QM}}^{\left(  \mathrm{geometric}\right)  }\leq1\Leftrightarrow
T_{AB}\geq\frac{\hslash}{\Delta E}\cos^{-1}\left[  \left\vert \left\langle
A|B\right\rangle \right\vert \right]  \label{cc1}%
\end{equation}
that is,%
\begin{equation}
\Delta E\cdot T_{AB}\geq\hslash\cos^{-1}\left[  \left\vert \left\langle
A|B\right\rangle \right\vert \right]  \text{.} \label{c2}%
\end{equation}
Eq. (\ref{c2}) generalizes the inequality $\Delta E\cdot T_{AB}^{\perp}\geq
h/4$ in Eq. (\ref{c1}) and is valid more generally even when the quantum
system does not pass through orthogonal states. Moreover, the inequality in
Eq. (\ref{c2}) is in agreement with Eq. (\ref{du10}) derived in the previous
Section without any geometrical consideration. The derivation of Eq.
(\ref{c2}) provides a simple quantitative justification of the verbal
statement made by Anandan and Aharonov in Ref. \cite{anandan90} concerning the
validity of the inequality $\varepsilon\leq1$ extended to a system that does
not pass through orthogonal states. In particular, we have from Eq. (\ref{c2})
that%
\begin{equation}
\text{geodesic motion}\Leftrightarrow\eta_{\text{QM}}^{\left(
\mathrm{geometric}\right)  }=1\Leftrightarrow\frac{\hslash}{\Delta E\cdot
T_{AB}}\cos^{-1}\left[  \left\vert \left\langle A|B\right\rangle \right\vert
\right]  =1\Leftrightarrow\Delta E\cdot T_{AB}=\hslash\cos^{-1}\left[
\left\vert \left\langle A|B\right\rangle \right\vert \right]  \text{,}%
\end{equation}
that is,%
\begin{equation}
\mathrm{geodesic\ motion}\Leftrightarrow\Delta E\cdot T_{AB}=\hslash\cos
^{-1}\left[  \left\vert \left\langle A|B\right\rangle \right\vert \right]
\text{.} \label{geo2}%
\end{equation}
We emphasize that the inequality $\eta_{\text{QM}}^{\left(  \mathrm{geometric}%
\right)  }\leq1$ in Eq. (\ref{cc1}) holds true also in the time-dependent
Hamiltonian scenario by simply applying our line of reasoning developed for
the time-independent scenario and replacing $\Delta E$ with $\left\langle
\Delta E\right\rangle _{T_{AB}}$ as defined in Eq. (\ref{minimi1}).
Specifically, the time-dependent version of Eq. (\ref{geo2}) for non-geodesic
motion is specified by the inequality $\left\langle \Delta E\right\rangle
_{T_{AB}}T_{AB}\geq\hslash\cos^{-1}\left[  \left\vert \left\langle
A|B\right\rangle \right\vert \right]  $.

Interestingly, we also point out that the energy dispersion $\Delta E$ (as
defined in Eq. (\ref{dispersion})) of a constant Hamiltonian operator
$\mathrm{\hat{H}}$ describing a two-levels quantum system with spectral
decomposition given by $\mathrm{\hat{H}}\overset{\text{def}}{=}E_{1}\left\vert
E_{1}\right\rangle \left\langle E_{1}\right\vert +E_{2}\left\vert
E_{2}\right\rangle \left\langle E_{2}\right\vert $ (where $E_{2}\geq E_{1}$
and $\left\langle E_{i}|E_{j}\right\rangle =\delta_{ij}$) with respect to the
normalized initial state $\left\vert A\right\rangle $ is given by,%
\begin{equation}
\Delta E=\frac{E_{2}-E_{1}}{2}\sqrt{1-\left(  \left\vert \alpha_{1}\right\vert
^{2}-\left\vert \alpha_{2}\right\vert ^{2}\right)  ^{2}}\text{,}\label{chi4}%
\end{equation}
once $\left\vert A\right\rangle $ is decomposed as $\alpha_{1}\left\vert
E_{1}\right\rangle +\alpha_{2}\left\vert E_{2}\right\rangle $ with $\alpha
_{1}$, $\alpha_{2}\in%
\mathbb{C}
$. From Eq. (\ref{chi4}), we note that the maximum value of $\Delta E$ is
obtained for $\left\vert \alpha_{1}\right\vert =\left\vert \alpha
_{2}\right\vert $ where $\left\vert \alpha_{1}\right\vert \overset{\text{def}%
}{=}\left\vert \left\langle E_{1}|A\right\rangle \right\vert $ and $\left\vert
\alpha_{2}\right\vert \overset{\text{def}}{=}\left\vert \left\langle
E_{2}|A\right\rangle \right\vert $, respectively. Moreover, this maximum value
equals $\Delta E_{\max}\overset{\text{def}}{=}\left(  E_{2}-E_{1}\right)  /2$.
Therefore, the minimum evolution time $T_{AB}$ from an initial state
$\left\vert A\right\rangle $ to a final state $\left\vert B\right\rangle $
becomes%
\begin{equation}
T_{AB}^{\min}=\frac{2\hslash}{E_{2}-E_{1}}\cos^{-1}\left[  \left\vert
\left\langle A|B\right\rangle \right\vert \right]  \text{.}\label{yields}%
\end{equation}
Finally, when the quantum evolution is between orthogonal\textbf{ }initial and
final states $\left\vert A\right\rangle $ and $\left\vert B\right\rangle $,
Eq. (\ref{yields}) yields $T_{AB}^{\perp\text{,}\min}=h/\left[  2\left(
E_{2}-E_{1}\right)  \right]  $. Clearly, from Eqs. (\ref{geo2}) and
(\ref{yields}) we observe that the travel time $T_{AB}$ depends on the
Hamiltonian \textrm{H} through the energy uncertainty $\Delta E$.
Specifically, $T_{AB}$ can be made arbitrarily small if $\Delta E$ can be made
arbitrarily large. However, in typical physical scenarios specified by a
finite-dimensional Hilbert space with temporally bounded energy eigenvalues
$\left\{  E_{n}\left(  t\right)  \right\}  $ \cite{ali09}, the dispersion of
the Hamiltonian operator is upper bounded. Specifically, it happens that
$\Delta E\leq\mathcal{E}_{\max}$ if for any $n\in%
\mathbb{N}
$ and for any $t$, one imposes $\left\vert E_{n}\left(  t\right)  \right\vert
\leq\mathcal{E}_{\max}$ for some $\mathcal{E}_{\max}\in%
\mathbb{R}
_{+}$. Thus, the minimum time travel is lower bounded with $T_{AB}^{\min}%
\geq\left(  \hslash/\mathcal{E}_{\max}\right)  \cos^{-1}\left[  \left\vert
\left\langle A|B\right\rangle \right\vert \right]  $. In Table I we summarize
our results for optimal quantum evolution conditions between both orthogonal,
and nonorthogonal states. In the next section, we present two explicit
examples. \begin{table}[t]
\centering
\begin{tabular}
[c]{c|c|c}\hline\hline
\textbf{Quantum States} & \textbf{Time-Energy Inequality Constraint} &
\textbf{Optimal Quantum Evolution Condition}\\\hline
orthogonal & $T_{AB}^{\perp}\geq h/(4\Delta E)$, with $0\leq\eta\leq1$ &
$T_{AB}^{\perp,\min}=h/(4\Delta E_{\max})$, with $\eta=1$\\
nonorthogonal & $T_{AB}\geq\hslash\cos^{-1}\left[  \left\vert \left\langle
A|B\right\rangle \right\vert \right]  /\Delta E$, with $0\leq\eta\leq1$ &
$T_{AB}^{\min}=\hslash\cos^{-1}\left[  \left\vert \left\langle
A|B\right\rangle \right\vert \right]  /\Delta E_{\max}$, with $\eta=1$\\\hline
\end{tabular}
\caption{Schematic description of the optimal quantum evolution conditions for
motion between either two orthogonal or nonorthogonal states. In both
scenarios, the energy dispersion $\Delta E^{2}$ of the Hamiltonian operator
\textrm{H }is assumed to be constant in time. Unit geometric quantum
efficiency motion with $\eta\overset{\text{def}}{=}\eta_{\text{QM}}^{\left(
\mathrm{geometric}\right)  }=1$ is achieved only when the time-energy
inequality relation becomes an equality constraint.}%
\end{table}

\section{Applications}

In this section, we explicitly discuss the notions of minimum evolution time
and quantum geometric efficiency in two examples.

\subsection{Time-independent scenario}

In the first scenario, we consider a physical system characterized by a
time-independent Hamiltonian
\begin{equation}
\mathrm{H}\overset{\text{def}}{=}\epsilon\sigma_{x}=\left(
\begin{array}
[c]{cc}%
0 & \epsilon\\
\epsilon & 0
\end{array}
\right)  \text{,} \label{hamo1}%
\end{equation}
with $\epsilon>0$ denoting the strength of the Hamiltonian and $\sigma_{x}$
being the usual Pauli matrix. The unitary evolution operator $U\left(
t\right)  \overset{\text{def}}{=}e^{-\frac{i}{\hslash}\mathrm{H}t}$
corresponding to the Hamiltonian in Eq. (\ref{hamo1}) is given by,%
\begin{equation}
U\left(  t\right)  =\cos\left(  \frac{\epsilon}{\hslash}t\right)
I-i\sin\left(  \frac{\epsilon}{\hslash}t\right)  \sigma_{x}=\left(
\begin{array}
[c]{cc}%
\cos\left(  \frac{\epsilon}{\hslash}t\right)  & -i\sin\left(  \frac{\epsilon
}{\hslash}t\right) \\
-i\sin\left(  \frac{\epsilon}{\hslash}t\right)  & \cos\left(  \frac{\epsilon
}{\hslash}t\right)
\end{array}
\right)  \text{,} \label{uni1}%
\end{equation}
with $I$ being the $2\times2$ identity matrix. Let us consider the quantum
mechanical evolution from $\left\vert A\right\rangle \overset{\text{def}}%
{=}\left\vert 0\right\rangle $ to $\left\vert B\right\rangle \overset
{\text{def}}{=}-i\left\vert 1\right\rangle $ along the path $\gamma
_{t}:t\mapsto\left\vert \psi\left(  t\right)  \right\rangle $ with $\left\vert
\psi\left(  t\right)  \right\rangle \overset{\text{def}}{=}U\left(  t\right)
\left\vert A\right\rangle $ and $0\leq t\leq T_{AB}^{\left(
\text{\textrm{effective}}\right)  }$ with $T_{AB}^{\left(
\text{\textrm{effective}}\right)  }\overset{\text{def}}{=}\frac{\pi}%
{2\epsilon}\hslash$. Note that $\left\vert A\right\rangle \overset{\text{def}%
}{=}\left\vert \psi\left(  0\right)  \right\rangle =\left\vert 0\right\rangle
$ and $\left\vert B\right\rangle \overset{\text{def}}{=}\left\vert \psi\left(
\frac{\pi}{2\epsilon}\hslash\right)  \right\rangle =-i\left\vert
1\right\rangle $ is physically equivalent to $\left\vert 1\right\rangle $.

To begin, we observe that the path $\gamma_{t}$ traced by the state vector
$\left\vert \psi\left(  t\right)  \right\rangle =\cos\left(  \frac{\epsilon
}{\hslash}t\right)  \left\vert A\right\rangle +\sin\left(  \frac{\epsilon
}{\hslash}t\right)  \left\vert B\right\rangle $ is a geodesic path. Indeed, a
simple calculation shows that $\left\vert \psi\left(  t\right)  \right\rangle
$ can be recast as a quantum geodesic line \cite{laba17},%

\begin{equation}
\left\vert \psi\left(  t\right)  \right\rangle =\left\vert \tilde{\psi}\left(
\xi\left(  t\right)  \right)  \right\rangle \overset{\text{def}}{=}%
\mathcal{N}_{\xi}\left[  \left(  1-\xi\right)  \left\vert A\right\rangle
+\xi\left\vert B\right\rangle \right]  \text{,}%
\end{equation}
where $\mathcal{N}_{\xi}\overset{\text{def}}{=}\left[  1-2\xi\left(
1-\xi\right)  \right]  ^{-1/2}$ is a normalization constant, while $\xi\left(
t\right)  $ denotes a strictly monotonic function of $t$ given by%
\begin{equation}
\xi\left(  t\right)  \overset{\text{def}}{=}\frac{\tan\left(  \frac{\epsilon
}{\hslash}t\right)  }{1+\tan\left(  \frac{\epsilon}{\hslash}t\right)
}\text{,}%
\end{equation}
with $0\leq\xi\left(  t\right)  \leq1$. The geodesic nature of the path
$\gamma_{t}$ can also be explained in terms of the energy spread and the
efficiency concepts. Indeed, we have $\Delta E^{2}=\epsilon^{2}$,
$T_{AB}^{\left(  \text{\textrm{effective}}\right)  }=\frac{\pi}{2\epsilon
}\hslash$, and $\left\langle A|B\right\rangle =0$. Therefore, the minimum
evolution constraint condition $\Delta E\cdot T_{AB}^{\left(
\text{\textrm{ideal}}\right)  }=h/4$ yields
\begin{equation}
T_{AB}^{\left(  \text{\textrm{effective}}\right)  }=T_{AB}^{\left(
\text{\textrm{ideal}}\right)  }\text{.}%
\end{equation}
Finally, the quantum motion occurs with unit geometric efficiency
$\eta_{\text{QM}}^{\left(  \mathrm{geometric}\right)  }=1$.

\subsection{Time-dependent scenario}

In the second scenario, we take into consideration a physical system specified
by means of a time-dependent Hamiltonian,%
\begin{equation}
\mathrm{H}\left(  t\right)  \overset{\text{def}}{=}\epsilon\cos\left(  \omega
t\right)  \sigma_{x}+\epsilon\sin\left(  \omega t\right)  \sigma_{y}%
+\frac{\hslash\omega_{0}}{2}\sigma_{z}=\left(
\begin{array}
[c]{cc}%
\frac{\hslash\omega_{0}}{2} & \epsilon e^{-i\omega t}\\
\epsilon e^{i\omega t} & -\frac{\hslash\omega_{0}}{2}%
\end{array}
\right)  \text{,} \label{hami2}%
\end{equation}
with $\epsilon>0$, $\omega>0$, and $\hslash\omega_{0}>0$ denoting the strength
of the external drive, the angular frequency of the external drive, and the
energy difference between the two states of the two-state quantum system (that
is, $\hslash\omega_{0}\overset{\text{def}}{=}E_{2}-E_{1}>0$). We define the
detuning of the driving field from resonance as\textbf{ }$\Delta
\overset{\text{def}}{=}\hbar\,(\omega-\omega_{0})$. Clearly, $\sigma_{x}$,
$\sigma_{y}$, and $\sigma_{z}$ are the usual Pauli matrices. The Hamiltonian
in Eq. (\ref{hami2}) emerges in the context of the near-resonance phenomenon
in a two--state quantum system. The unitary evolution operator $U\left(
t\right)  $ corresponding to $\mathrm{H}\left(  t\right)  $ in Eq.
(\ref{hami2}) is \cite{scully},%
\begin{align}
U\left(  t\right)   &  =\cos\left(  \frac{\kappa}{\hslash}t\right)
I-i\sin\left(  \frac{\kappa}{\hslash}t\right)  \left[  \frac{\Delta}{2\kappa
}\sigma_{z}+\frac{\epsilon}{\kappa}\sigma_{x}\right] \nonumber\\
& \\
&  =\left(
\begin{array}
[c]{cc}%
\cos\left(  \frac{\kappa}{\hslash}t\right)  -i\frac{\Delta}{2\kappa}%
\sin\left(  \frac{\kappa}{\hslash}t\right)  & -i\frac{\epsilon}{\kappa}%
\sin\left(  \frac{\kappa}{\hslash}t\right) \\
-i\frac{\epsilon}{\kappa}\sin\left(  \frac{\kappa}{\hslash}t\right)  &
\cos\left(  \frac{\kappa}{\hslash}t\right)  +i\frac{\Delta}{2\kappa}%
\sin\left(  \frac{\kappa}{\hslash}t\right)
\end{array}
\right)  \text{,}\nonumber
\end{align}
where $\kappa\overset{\text{def}}{=}\sqrt{\epsilon^{2}+\Delta^{2}/4}$. Let us
consider the quantum mechanical evolution from $\left\vert A\right\rangle
\overset{\text{def}}{=}\left\vert 0\right\rangle $ to $\left\vert
B\right\rangle \overset{\text{def}}{=}-i\frac{\Delta}{2\kappa}\left\vert
0\right\rangle -i\frac{\epsilon}{\kappa}\left\vert 1\right\rangle $ with
$0\leq t\leq T_{AB}^{\left(  \text{\textrm{effective}}\right)  }$ where
$T_{AB}^{\left(  \text{\textrm{effective}}\right)  }\overset{\text{def}}%
{=}\frac{\pi}{2\kappa}\hslash$. As a side remark, we note that near-resonance,
$\left\vert \Delta\right\vert \ll\epsilon$ and $\kappa\rightarrow\epsilon$.
Therefore, $\frac{\pi}{2\kappa}\hslash\rightarrow\frac{\pi}{2\epsilon}%
\hslash\overset{\text{def}}{=}\left[  T_{AB}^{\left(  \text{effective}\right)
}\right]  _{\text{first-scenario}}$, and $-i\frac{\Delta}{2\kappa}\left\vert
0\right\rangle -i\frac{\epsilon}{\kappa}\left\vert 1\right\rangle
\rightarrow-i\left\vert 1\right\rangle \overset{\text{def}}{=}\left[
\left\vert B\right\rangle \right]  _{\text{first-scenario}}$.

To begin, we note that the path $\gamma_{t}$ traced by the state vector
$\left\vert \psi\left(  t\right)  \right\rangle =\cos\left(  \frac{\kappa
}{\hslash}t\right)  \left\vert A\right\rangle +\sin\left(  \frac{\kappa
}{\hslash}t\right)  \left\vert B\right\rangle $ with $\left\langle
A|B\right\rangle =-i\frac{\Delta}{2\kappa}\neq0$ is not a geodesic path.
Indeed, following the reasoning outlined in the first example, a simple
calculation shows that $\left\vert \psi\left(  t\right)  \right\rangle $
cannot be recast as a quantum geodesic line. The non-geodesic nature of the
path $\gamma_{t}$ can also be understood by studying the expression of the
energy spread of the system. Indeed, from a simple calculation, we obtain%
\begin{equation}
\Delta E^{2}\left(  t\right)  =\varepsilon^{2}+\frac{\left(  \hslash\omega
_{0}\right)  ^{2}}{4}-\left\{
\begin{array}
[c]{c}%
\frac{\hslash\omega_{0}}{2}\left[  \cos^{2}\left(  \frac{\kappa}{\hslash
}t\right)  -\frac{4\varepsilon^{2}-\Delta^{2}}{4\kappa^{2}}\sin^{2}\left(
\frac{\kappa}{\hslash}t\right)  \right]  -\frac{\varepsilon^{2}}{\kappa}%
\sin\left(  2\frac{\kappa}{\hslash}t\right)  \sin\left(  \omega t\right)  +\\
+\frac{2\varepsilon^{2}}{\kappa}\frac{\Delta}{2\kappa}\sin^{2}\left(
\frac{\kappa}{\hslash}t\right)  \cos\left(  \omega t\right)
\end{array}
\right\}  ^{2}\text{.} \label{deltaenergy}%
\end{equation}
At this point, to show the non-geodesic behavior of the evolution path
$\gamma_{t}$ in this second scenario, we focus on the short-time limit of
$\Delta E^{2}\left(  t\right)  $ in the near-resonance case. First, assuming
$\left\vert \Delta\right\vert \ll\epsilon$, $\Delta E^{2}\left(  t\right)  $
in Eq. (\ref{deltaenergy}) reduces to%
\begin{equation}
\Delta E^{2}\left(  t\right)  =\varepsilon^{2}+\frac{\left(  \hslash\omega
_{0}\right)  ^{2}}{4}\left\{  1-\left[  \cos\left(  2\frac{\varepsilon
}{\hslash}t\right)  -\frac{2\varepsilon}{\hslash\omega_{0}}\sin\left(
2\frac{\varepsilon}{\hslash}t\right)  \sin\left(  \omega t\right)  \right]
^{2}\right\}  \text{,} \label{delta2}%
\end{equation}
with $0\leq t\leq\frac{\pi}{2\epsilon}\hslash$. Before considering the
short-time limit, we make a few considerations that motivate the consideration
of this limit. The strength of the external drive $\epsilon$ is related to the
Rabi angular frequency by the relation $\epsilon=\hslash\Omega_{\mathrm{Rabi}%
}$ \cite{sakurai}, with $\Omega_{\mathrm{Rabi}}\overset{\text{def}}{=}%
eB_{\bot}/(2mc)$. The quantities $e$ and $m$ are the charge and the mass of an
electron. The quantity $c$ denotes the speed of light, while $B_{\bot}$ is the
intensity of the magnetic field originating from the magnetic field components
that are in the plane orthogonal to the quantization axis (that is, the
$z$-axis). The energy gap $\hslash\omega_{0}$ is related to the Larmor angular
frequency, $\hslash\omega_{0}=\hslash\Omega_{\mathrm{Larmor}}$ \cite{sakurai},
with $\Omega_{\mathrm{Larmor}}\overset{\text{def}}{=}eB_{\parallel}/(mc)$. The
quantity $B_{\parallel}$ denotes the intensity of the magnetic field along the
quantization axis. The Larmor frequency $\nu_{_{\mathrm{Larmor}}}%
\overset{\text{def}}{=}\Omega_{\mathrm{Larmor}}/(2\pi)$ of an electron in a
magnetic field with $B_{\parallel}=1\mathrm{T}$ is,%
\begin{equation}
\nu_{_{\mathrm{Larmor}}}=\frac{1}{2\pi}\left(  \frac{eB_{\parallel}}%
{mc}\right)  _{\mathrm{cgs}}=\frac{1}{2\pi}\left(  \frac{eB_{\parallel}}%
{m}\right)  _{\mathrm{MKSA}}\simeq28\mathrm{GHz}\text{,}%
\end{equation}
with \textquotedblleft\textrm{cgs}\textquotedblright\ and \textquotedblleft%
\textrm{MKSA}\textquotedblright\ denote the physical unit system being used.
Magnetic field intensities $B_{\parallel}$ employed in MRI (magnetic resonance
imaging) are typically in the range of $1$ to $4$ $\mathrm{T}$. In the
weak-driving regime \cite{shim14}, consider a typical scenario where
$B_{\parallel}\gg B_{\bot}\simeq10^{-6}\mathrm{T}=10^{-2}\mathrm{G}$ and
$T_{AB}^{\left(  \text{\textrm{effective}}\right)  }=\frac{\pi}{2\epsilon
}\hslash\simeq1.8\times10^{-5}\mathrm{\sec.}\ll1$. Observe that since
$T_{AB}^{\left(  \text{\textrm{effective}}\right)  }\propto1/B_{\bot}$, the
condition $T_{AB}^{\left(  \text{effective}\right)  }\ll1$ is satisfied by
larger values of $B_{\bot}$ as well. Therefore, assuming to be in the
short-time limit, from $\Delta E^{2}\left(  t\right)  $ in Eq. (\ref{delta2})
we Taylor expand $\Delta E\left(  t\right)  $ in the neighborhood of
$t=0$.\ Then, the energy spread $\Delta E\left(  t\right)  $ reduces to%
\begin{equation}
\Delta E\left(  t\right)  =\varepsilon\left[  1+\frac{\omega_{0}^{2}}%
{2}\left(  1+\frac{2\omega}{\omega_{0}}\right)  t^{2}\right]  +\mathcal{O}%
\left(  t^{4}\right)  \text{,} \label{delta3}%
\end{equation}
with $\mathcal{O}\left(  t^{4}\right)  $ denoting an infinitesimal quantity of
order four or higher\textbf{. }From the condition that defines $T_{AB}%
^{\left(  \text{\textrm{ideal}}\right)  }$,%
\begin{equation}
\int_{0}^{T_{AB}^{\left(  \text{\textrm{ideal}}\right)  }}\Delta E\left(
t\right)  dt=\hslash\frac{\pi}{2}\text{,}%
\end{equation}
we get using Eq. (\ref{delta3}),%
\begin{equation}
T_{AB}^{\left(  \text{\textrm{effective}}\right)  }=\frac{\pi}{2\epsilon
}\hslash=\int_{0}^{T_{AB}^{\left(  \text{\textrm{ideal}}\right)  }}\left[
1+\frac{\omega_{0}^{2}}{2}\left(  1+\frac{2\omega}{\omega_{0}}\right)
t^{2}\right]  dt>\int_{0}^{T_{AB}^{\left(  \text{\textrm{ideal}}\right)  }%
}dt=T_{AB}^{\left(  \text{\textrm{ideal}}\right)  }\text{.}
\label{inequality1}%
\end{equation}
The above inequality is justified by the positivity of the second addendum in
the integrand in Eq. (\ref{inequality1}). Therefore, $T_{AB}^{\left(
\text{\textrm{ideal}}\right)  }$ is upper bounded by the effective evolution
time $T_{AB}^{\left(  \text{\textrm{effective}}\right)  }$. In particular,
calculating the integral in Eq. (\ref{inequality1}), $T_{AB}^{\left(
\text{\textrm{ideal}}\right)  }\equiv\tilde{T}$ is implicitly defined by the
relation%
\begin{equation}
\frac{\pi}{2\epsilon}\hslash=\tilde{T}+\frac{1}{3}a\tilde{T}^{3}\text{,}
\label{implicit}%
\end{equation}
with $a\overset{\text{def}}{=}\frac{\omega_{0}^{2}}{2}\left(  1+\frac{2\omega
}{\omega_{0}}\right)  >0$ and $\left[  a\right]  _{\mathrm{MKSA}}%
=\mathrm{\sec}$.$^{-2}$. Clearly, Eq. (\ref{implicit}) is a special case of
Eq. (\ref{minimi}) and can only be solved numerically for $\tilde{T}$ once
$\epsilon$, $\omega$, and $\omega_{0}$ are fixed. From Eq. (\ref{inequality1}%
), we expect to get $\tilde{T}<T_{AB}^{\left(  \text{\textrm{effective}%
}\right)  }$ with the quantum motion occurring with geometric efficiency
$\eta_{\text{QM}}^{\left(  \mathrm{geometric}\right)  }<1$.

The discussion of this time-dependent scenario illustrates the type of
challenges that one may encounter when dealing with more realistic
time-dependent scenarios.

\section{Concluding remarks}

In this paper, we presented a simple proof of the fact that the minimum time
$T_{AB}$ for the quantum evolution between two arbitrary states $\left\vert
A\right\rangle $ and $\left\vert B\right\rangle $ equals $T_{AB}=\hslash
\cos^{-1}\left[  \left\vert \left\langle A|B\right\rangle \right\vert \right]
/\Delta E$ (see Eq.\ (\ref{du10})) with $\Delta E$ being the constant energy
uncertainty of the system. This proof was performed in the absence of any
geometrical arguments and followed closely the reasoning employed in Ref.
\cite{vaidman92} by Vaidman. Then, within the geometric framework of quantum
evolutions based upon the geometry of the projective Hilbert space as
developed by Anandan and Aharonov in Ref. \cite{anandan90}, we discussed the
roles played by either minimum-time or maximum-energy uncertainty concepts in
defining a geometric efficiency measure (see $\eta_{\text{QM}}^{\left(
\mathrm{geometric}\right)  }$ in Eq. (\ref{effi})) of quantum evolutions
between two arbitrary nonorthogonal quantum states. In particular, we provided
a quantitative justification of the validity of the inequality $\eta
_{\text{QM}}^{\left(  \mathrm{geometric}\right)  }\leq1$ even when the system
passed through nonorthogonal states (see Eq. (\ref{c2})). A schematic
description of our main discussion points appears in Table I.

While our investigation is performed in the spirit of the original Vaidman
work, we additionally consider here a number of new modifications. Firstly, we
extend the reasoning to unitary Schr\"{o}dinger evolutions between quantum
states that are not necessarily orthogonal. Secondly, we provide two explicit
and alternative detailed proofs of the clever decomposition of $\hat
{Q}\left\vert \psi\right\rangle $ in Eq. (\ref{deco}) which plays a key role
in the main proof itself. Thirdly, we emphasize its generalization to
time-dependent Hamiltonian evolutions. Fourthly, and perhaps most importantly,
we show the usefulness of the outcomes of the proof in upper bounding the
geometric efficiency of quantum evolutions between two arbitrary states,
either orthogonal or nonorthogonal. Lastly, we quantitatively present two
illustrative examples discussing both time-independent and time-dependent
quantum Hamiltonian evolutions in terms of minimum evolution time and
geometric efficiency. 

As a final remark, we point out that it would be interesting to further deepen
our understanding of this geometric efficiency analysis to physical scenarios
where the energy uncertainty $\Delta E$ is not constant in time. A partial
list of scenarios that could be considered includes the\textbf{ }%
$\mathrm{su}\left(  2\text{; }%
\mathbb{C}
\right)  $ time-dependent\textbf{ }Hamiltonian evolutions used to describe
distinct types of analog quantum search schemes viewed as driving strategies
in Ref. \cite{cafaro19} and, in addition, the time-dependent Hamiltonian
describing the resonance phenomenon in a two-state quantum system used to
construct quantum search algorithms by Wilczek and collaborators in Ref.
\cite{frank20} without limiting the analysis to the short-time limit of the
near-resonance regime. We hope to address these more applied investigations in
future efforts.

\begin{acknowledgments}
C. C. is grateful to the United States Air Force Research Laboratory (AFRL)
Summer Faculty Fellowship Program for providing support for this work. P. M.
A. acknowledges support from the Air Force Office of Scientific Research
(AFOSR). Any opinions, findings and conclusions or recommendations expressed
in this material are those of the author(s) and do not necessarily reflect the
views of the Air Force Research Laboratory (AFRL).
\end{acknowledgments}

\appendix

\section{Alternative derivation of Eq. (\ref{deco})}

In this Appendix, we provide an alternative derivation of Eq. (\ref{deco}).
Let us assume that the full Hilbert space $\mathcal{H}$ has an orthogonal
decomposition given by $\mathcal{H}\overset{\text{def}}{=}\mathcal{H}_{\psi
}\oplus\mathcal{H}_{\psi_{\bot}}$. Therefore, the identity operator
$\mathrm{\hat{1}}$ on $\mathcal{H}$ can be decomposed in terms of orthogonal
projector operators as $\mathrm{\hat{1}}\overset{\text{def}}{=}P_{\left\vert
\psi\right\rangle }+P_{\left\vert \psi_{\bot}\right\rangle }$ with
$P_{\left\vert \psi\right\rangle }$ and $P_{\left\vert \psi_{\bot
}\right\rangle }$ given by $\left\vert \psi\right\rangle \left\langle
\psi\right\vert $ and $\left\vert \psi_{\bot}\right\rangle \left\langle
\psi_{\perp}\right\vert $, respectively. Then, for any operator $\hat{Q}$, the
state $\hat{Q}\left\vert \psi\right\rangle $ can be recast as%
\begin{align}
\hat{Q}\left\vert \psi\right\rangle  &  =\mathrm{\hat{1}}\hat{Q}\left\vert
\psi\right\rangle \nonumber\\
&  =\left(  P_{\left\vert \psi\right\rangle }+P_{\left\vert \psi_{\bot
}\right\rangle }\right)  \hat{Q}\left\vert \psi\right\rangle \nonumber\\
&  =\left\langle \psi\right\vert \hat{Q}\left\vert \psi\right\rangle
\left\vert \psi\right\rangle +P_{\left\vert \psi_{\bot}\right\rangle }\hat
{Q}\left\vert \psi\right\rangle \text{,}%
\end{align}
that is, setting $\left\langle \hat{Q}\right\rangle \overset{\text{def}}%
{=}\left\langle \psi\right\vert \hat{Q}\left\vert \psi\right\rangle $,
\begin{equation}
\hat{Q}\left\vert \psi\right\rangle =\left\langle \hat{Q}\right\rangle
\left\vert \psi\right\rangle +\left[  \left\langle \psi|\left(  P_{\left\vert
\psi_{\bot}\right\rangle }\hat{Q}\right)  ^{\dagger}\left(  P_{\left\vert
\psi_{\bot}\right\rangle }\hat{Q}\right)  |\psi\right\rangle \right]
^{1/2}\frac{P_{\left\vert \psi_{\bot}\right\rangle }\hat{Q}\left\vert
\psi\right\rangle }{\left[  \left\langle \psi|\left(  P_{\left\vert \psi
_{\bot}\right\rangle }\hat{Q}\right)  ^{\dagger}\left(  P_{\left\vert
\psi_{\bot}\right\rangle }\hat{Q}\right)  |\psi\right\rangle \right]  ^{1/2}%
}\text{.} \label{peppe}%
\end{equation}
Assuming that $\hat{Q}$ is an Hermitian operator, it happens that the
amplitude $\left\langle \psi|\left(  P_{\left\vert \psi_{\bot}\right\rangle
}\hat{Q}\right)  ^{\dagger}\left(  P_{\left\vert \psi_{\bot}\right\rangle
}\hat{Q}\right)  |\psi\right\rangle $ equals the dispersion $\Delta Q^{2}$ of
the operator $\hat{Q}$ defined as $\Delta Q^{2}\overset{\text{def}}%
{=}\left\langle \psi|\hat{Q}^{2}|\psi\right\rangle -\left\langle \psi|\hat
{Q}|\psi\right\rangle ^{2}$. Therefore, Eq. (\ref{peppe}) becomes%
\begin{equation}
\hat{Q}\left\vert \psi\right\rangle =\left\langle \hat{Q}\right\rangle
\left\vert \psi\right\rangle +\Delta Q\left\vert \psi_{\bot}\right\rangle
\text{,} \label{deco11}%
\end{equation}
with the unit vector $\left\vert \psi_{\bot}\right\rangle $\textbf{ }in Eqs.
(\ref{deco11}) given by,%
\begin{equation}
\left\vert \psi_{\bot}\right\rangle \overset{\text{def}}{=}\frac{P_{\left\vert
\psi_{\bot}\right\rangle }\hat{Q}\left\vert \psi\right\rangle }{\Delta
Q}=\frac{\hat{Q}\left\vert \psi\right\rangle -\left\langle \hat{Q}%
\right\rangle \left\vert \psi\right\rangle }{\Delta Q}\text{.} \label{deco2}%
\end{equation}
{The expression in Eq.(\ref{deco2}) explicitly exhibits the orthonormal nature
of $|\psi\rangle$ and $|\psi_{\perp}\rangle$, namely $\langle\psi_{\perp}%
|\psi_{\perp}\rangle=1$ and $\langle\psi|\psi_{\perp}\rangle=0$. }

Eqs. (\ref{deco11}) and (\ref{deco2}) conclude our alternative proof of the
simple, yet important, formula in Eq. (\ref{deco}).

\end{document}